# Deciphering Dynamical Nonlinearities in Short Time Series Using Recurrent Neural Networks

Radhakrishnan Nagarajan

Division of Biomedical Informatics, University of Kentucky, 725 Rose Street, MDS 230F, Lexington, KY 40536-0082, USA, Email: rnagarajan@uky.edu

**ABSTRACT**
Surrogate testing techniques have been used widely to investigate the presence of dynamical nonlinearities, an essential ingredient of deterministic chaotic processes. Traditional surrogate testing subscribes to statistical hypothesis testing and investigates potential differences in discriminant statistics between the given empirical sample and its surrogate counterparts. The choice and estimation of the discriminant statistics can be challenging across short time series. Also, conclusion based on a single empirical sample is an inherent limitation. The present study proposes a recurrent neural network classification framework that uses the raw time series obviating the need for discriminant statistic while accommodating multiple time series realizations for enhanced generalizability of the findings. The results are demonstrated on short time series with lengths (L = 32, 64, 128) from continuous and discrete dynamical systems in chaotic regimes, nonlinear transform of linearly correlated noise and experimental data. Accuracy of the classifier is shown to be markedly higher than >> 50% for the processes in chaotic regimes whereas those of nonlinearly correlated noise were around ~50% similar to that of random guess from a one-sample binomial test. These results are promising and elucidate the usefulness of the proposed framework in identifying potential dynamical nonlinearities from short experimental time series.

**Introduction**
Time series data can be realized by discretizing a continuous process in amplitude and time. Discretization in amplitude is a result of quantization whereas discretization in time can be achieved using an optimal sampling frequency (e.g. Nyquist rate)[1] for certain class of processes. Understanding the correlation structure is fundamental to time series analysis and can provide critical insights into its generative mechanism. On a related note, optimal parameters of a linearly correlated processes such as auto-regressive process can be estimated faithfully from their auto-correlation function (Yule-Walker equations)[1]. Auto-correlation in turn is related to their power-spectral density representing the distribution of the power across the various frequencies by the Wiener-Khintchine theorem[1]. Parametric as well as non-parametric approaches have been used widely for spectral estimation. Of interest is to note that non-parametric approaches such as subspace decomposition (Pisarenko Harmonic Decomposition)[1] estimate the dominant frequencies by eigen-decomposition of the corresponding Toeplitz matrix whose elements are essentially the auto-correlation function. On the other hand, correlation signatures in a given time series need not necessarily be linear. Nonlinear correlations can arise as a result of static nonlinearities as well as dynamical nonlinearities. Static nonlinearities are often attributed to the transfer function of a measurement device (e.g. sensor) that maps an analog or continuous process onto digital data. In contrast, dynamical nonlinearities such as those from nonlinear deterministic systems are a result of nonlinear coupling and can exhibit a wide-range of intricate behaviors including deterministic chaos[2-8]. Identifying chaos can be helpful in developing suitable approaches for their control [9,10]. Chaos has also been shown to have a wide-range of applications[11]. Break down in dynamical nonlinearities have also been shown to discriminate health and disease[3]. It is important to note that spectral analysis while useful for investigating narrow-band processes can be singularly unhelpful in adequately describing chaotic processes as they exhibit a broad-band spectrum similar to that of noise[12]. On a related note, linear filtering used widely to minimize the effect of noise have been shown to introduce marked distortion of the phase-space geometry of time series from chaotic systems[13]. Takens embedding procedure[14,15] provided an elegant way to reconstruct the multi-dimensional phase-space representation of nonlinear dynamical systems from their univariate time series representation using an appropriate time delay and embedding dimension[15,16]. It was perhaps one of the primary drivers in investigating the presence of deterministic chaos from time series realizations. Subsequently, an array of approaches with ability to provide insight into the generative mechanism behind a given time series under the broad theme "surrogate testing" were proposed. Surrogate testing is similar to statistical resampling techniques[17] and used widely to investigate the presence of dynamical nonlinearities in experimental time series[18-25]. On a related note, dynamical nonlinearities are an essential ingredient of deterministic chaotic processes. There have been several noteworthy contributions to surrogate testing from the statistical physics community[26-33] summarized in recent reviews [27,34].

Essential ingredients of classical surrogate testing include (a) an empirical time series sample, (b) null hypothesis, (c) discriminant statistic or dynamical invariant, (d) surrogate generation algorithm and (e) a statistical test. The empirical sample has traditionally been a single time series realization from the given system of interest. The null hypothesis assumes the generative mechanism of the given empirical sample. Surrogate algorithms are designed to generate time series realizations (i.e. surrogates) from the given empirical sample retaining critical properties that align with the null hypothesis. For these reasons, surrogates are also regarded as constrained randomized realizations[27,35]. Several surrogate generation algorithms have been proposed in literature. These include (a) Random Shuffled Surrogates, (b) Phase-Randomized Surrogates (Fourier Transform Surrogates, FT)[26], (c) Amplitude Adjusted Fourier Transform Surrogates (AAFT) and (d) Iterated Amplitude Adjusted Fourier Transform (IAAFT)[26-28]. Each of these surrogate algorithms addresses a particular null hypothesis. Random shuffled surrogate investigates whether the given empirical sample is uncorrelated noise and retains the probability distribution of the empirical sample in the surrogate realization destroying the correlation in the empirical sample. Thus any discriminant statistic sensitive to the correlation in the given data can be used as a discriminant statistic. FT surrogates preserve the power-spectrum of the given empirical sample in the surrogate realizations by constrained randomization of the phases. As noted earlier, preserving the power-spectrum is sufficient to determine the optimal parameter of linearly correlated processes. FT surrogates can be used to investigate the presence of nonlinear correlation in the given empirical sample but does not provide insight into the nature of nonlinearity. Thus any discriminant statistic sensitive to nonlinear correlations is a reasonable choice for FT surrogates. Subsequently, AAFT surrogates[26] were proposed in order to address the null hypothesis that the given process is a static, invertible nonlinear transform of a linearly correlated noise by following a phase-randomization and rank ordering procedure. IAAFT surrogates[28] has been shown to preserve the spectrum as well as the probability distribution of the given empirical sample in the surrogate realization while overcoming the flatness bias prevalent in AAFT surrogates. The primary objective of IAAFT surrogates was to identify potential dynamical nonlinearities in the given time series. Thus any discriminant statistic sensitive to dynamical nonlinearities (e.g. dynamical invariants) can be used for AAFT and IAAFT surrogates. Several additional surrogate algorithms have also been proposed since then[34]. However, surrogates in the present study are generated using the IAAFT surrogates. Finally, parametric and non-parametric statistical tests were proposed to assess significant difference in the discriminant statistic estimates between the empirical sample and the surrogate counterparts[27].

Traditional surrogate testing approaches while helpful have inherent limitations. They primarily rely on statistical comparison of discriminant statistic estimates on a single representative sample (i.e. empirical sample) to those obtained on its corresponding surrogate realizations, **Fig. 1a**. While the choice of empirical sample can be attributed to implicit ergodic assumptions[36], generating long time series so as to enable robust estimation of dynamical invariants and discriminant statistics can be especially challenging in experimental settings as it demands controlling a number of factors. Experimental time series such as those from physiological systems have been especially known to exhibit variations between subjects within a given disease group or cohort. These in turn encourages accommodating multiple realizations as opposed to a single empirical sample in the surrogate testing framework for enhanced generalizability of the findings. In such a scenario, each realization can be paired with the corresponding surrogate realization, **Fig. 1b**. As in the case of single empirical sample, if the multiple time series realizations are sufficiently long then it might be possible to statistically compare the distribution of discriminant statistic estimates on the given cohort to those estimated on its paired surrogate realizations addressing the null hypothesis that there is no significant difference in the discriminant estimates between the cohort and its surrogate counterpart, **Fig. 1b**. The present study takes a different tack to the classical surrogate testing. Its significance can be attributed to the following reasons. (a) The present study proposes a binary classification framework that uses a simple recurrent neural network with the raw time series as the input obviating the need to choose or estimate discriminant statistics or dynamical invariants. This is especially helpful across small lengths such as those discussed in the present study (L = 32, 64, 128) where estimation of discriminant statistics[37] can be challenging and unreliable. (b) It poses the classical statistical surrogate testing **Figs. 1a-b**, as a binary classification problem, **Fig. 1c**, using recurrent neural networks (RNN), **Fig. 2**, where the two classes of interest correspond to the multiple time series realizations from a given cohort and their corresponding IAAFT surrogate counterparts. Generalizability of the proposed approach is established by demonstrating the classifier performance on an independent validation data. (c) The results are demonstrated on short time series of lengths (L = 32, 64, 128) generated by nonlinear deterministic processes in chaotic regimes, nonlinear transforms of linearly correlated noise with varying parameters as well as experimental time series data.

**Results**
Accuracy of the binary classification framework was investigated across nonlinear deterministic, experimental time series and nonlinear transform of linearly correlated noise (Sec. Methods) with lengths (L = 32, 64, 128), **Fig. 3**. Only length (L = 128) was considered for the epileptic seizure in order to faithfully represent at least a few cycles of the seizure dynamics. Convergence of RNN training and validation loss for representative time series realizations is shown in **Fig. 4**. Accuracy of the test data as a function of the epochs for each of these time series are shown in **Figs. 5-7** respectively. Representative accuracies for each of these data sets chosen from the plateau region of the plots where the training and validation loss were consistently low are enclosed in **Table 1**.

*Nonlinear Deterministic Process*
For time series generated from discrete and continuous nonlinear deterministic systems (Logistic, Henon, Lorenz and Rossler, Sec. Methods), the accuracy of the classifier showed a marked transition towards larger values from 0.5 as a function of the epochs, **Fig. 5**. A one-sample binomial test rejected the null that the accuracy was similar to that of random guess (0.5) at a significance level ($\alpha = 0.05$), **Table 1**. These results were consistently observed across the three sample sizes (L = 32, 64, 128) and across the data sets demonstrating the classifiers ability to discern dynamical nonlinearities and their IAAFT surrogate counterparts. The number of neurons in the hidden layer of the RNN was fixed at (N = 10). The RNN parameters (Sec. Methods) were fixed across these data sets, **Table 1**.

*Experimental Time Series*
Experimental time series generated using Chua's circuits (L = 32, 64, 128) and Santa Fe Laser Time Series (L = 32, 64, 128) in chaotic regimes (Sec. Methods) exhibited accuracies much greater than 0.5, **Fig. 6**, as observed in the case of the nonlinear deterministic processes, **Fig. 5**. A one-sample binomial test rejected the null hypothesis that the representative accuracy was similar to that of random guess (0.5) at a significance level ($\alpha = 0.05$), **Table 1**. For the time series generated from Chua's circuits and the Santa Fe laser time series, the number of neurons in the hidden layer of the RNN were chosen as 20 for (L = 32, 64) and 25 for (L = 128), **Table 1**. All other parameters of the RNN were retained as discussed in (Sec. Methods). Three representative EEG signals of lengths (L = 128) during seizure from a recent study[3] were reinvestigated using the proposed approach. Unlike Chua's circuits and Santa Fe time series, it is important to note that the underlying process generating the EEG signals during seizures is unknown. However, several studies have investigated nonlinear dynamical aspects of seizures and the evolution of characteristic synchronization patterns accompanying seizures[38,39]. The accuracy of the classifier as a function of the epoch exhibited a marked transition from 0.5 for the EEG. A one-sample binomial test rejected the null that the representative accuracy was similar to that of random guess (0.5) at a significance level (α = 0.05), **Table 1**. The number of neurons in the hidden layer of the RNN was fixed at (N = 20) for the three EEG signals, **Fig. 6**. All other parameters of the RNN were retained as discussed in (Sec. Methods).

**Table 1** Classification accuracies for nonlinear deterministic processes in chaotic regimes, experimental time series and non-deterministic processes. Accuracy estimates that were statistically significant (α = 0.05) from 0.5 in a one-sample binomial test are shown by asterisk.

| Time Series | N = 32 | N = 64 | N = 128 |
|---|---|---|---|
| *Nonlinear Deterministic Processes* | | | |
| *Logistic* (10 neurons) | 0.98* | 0.97* | 0.96* |
| *Henon* (10 neurons) | 0.93* | 0.96* | 0.87* |
| *Lorenz* (10 neurons) | 0.98* | 0.98* | 0.97* |
| *Rossler* (10 neurons) | 0.94* | 0.82* | 0.83* |
| *Experimental Time Series Data* | | | |
| *Santa Fe Laser Time Series* | 0.94* (20 neurons) | 0.84* (20 neurons) | 0.91* (25 neurons) |
| *Chua's Oscillator* | 0.82* (20 neurons) | 0.92* (20 neurons) | 0.98* (25 neurons) |
| *Epileptic Seizure 1* | | | 0.95* (20 neurons) |
| *Epileptic Seizure 2* | | | 0.95* (20 neurons) |
| *Epileptic Seizure 3* | | | 0.98* (20 neurons) |
| *Nonlinearly Correlated Noise* | | | |
| $\alpha = 0.2$ (10 neurons) | 0.53 | 0.48 | 0.52 |
| $\alpha = 0.4$ (10 neurons) | 0.49 | 0.50 | 0.51 |
| $\alpha = 0.6$ (10 neurons) | 0.51 | 0.46 | 0.49 |
| $\alpha = 0.8$ (10 neurons) | 0.49 | 0.53 | 0.50 |

*Nonlinear Transform of Linearly Correlated Noise*
Time series generated from a static nonlinear transform of linearly correlated noise[28] (Sec. Methods) were investigated with varying process parameters ($\alpha = 0.2, 0.4, 0.6, 0.8$) in the stationary regime, **Fig. 7**. Unlike the case of nonlinear deterministic chaos, accuracy estimates from the RNN classification framework did not show an appreciable change from that of random guess (0.5), **Fig. 7** as expected, indicating that the properties of the given data are not significantly different from those of their IAAFT surrogate counterparts. A one-sample binomial test did not reject the null that the representative accuracy was similar to that of random guess (0.5) at a significance level (α = 0.05), **Table 1**. These results were consistent across the different process parameters ($\alpha = 0.2, 0.4, 0.6, 0.8$) and lengths (L = 32, 64, 128). The number of neurons in the hidden layer of the RNN were fixed at (N = 10) similar to that of the nonlinear deterministic processes. Any further increase in the number of neurons in the hidden layer resulted in overfitting like behavior accompanied by marked separation in the training and validation loss. All other parameters of the RNN were retained as discussed in (Sec. Methods).

**Discussion**
Several studies have successfully used surrogate testing techniques to discern static and dynamical nonlinearities such as those from deterministic chaotic systems. Their ability to provide insights into the generative mechanism from the given time series realization(s) is a primary reason for their widespread adoption across a spectrum of disciplines. Traditional surrogate testing while helpful has inherent limitations. It subscribes to statistical hypothesis testing and investigates the separation of a chosen discriminant statistic or dynamical invariant between the given empirical sample and its surrogate counterpart. These discriminant statistic and dynamical invariants essentially capture certain facets of the given time series and their choice can be non-trivial with marked impact on the conclusions. Dynamical invariants and discriminant statistic estimation can be especially challenging across short time series such as those discussed in the present study. The proposed approach obviates the need to estimate discriminant statistics or dynamical invariants and uses the raw time series in the surrogate testing procedure. Conclusions based on traditional surrogate testing are also based on single realization or empirical sample. However, drawing conclusions based on a single realization can be a limitation from a practical standpoint. This is especially true with experimental data such as those from physiological systems and healthcare settings where variations are common within a given cohort. These in turn demand incorporation of multiple realizations for enhanced generalizability with potential to assist in clinical decision making. The proposed approach accommodates multiple realizations simultaneously and poses the traditional statistical hypothesis testing framework as a classification framework. For the nonlinear deterministic process, a marked increase in accuracy was observed as a function of epochs unlike that of the non-deterministic processes. Ideally, the error rate (i.e. 1 – accuracy) distribution may be positively skewed for large number of epochs for the nonlinear deterministic whereas that of non-deterministic process is expected to be relatively uniform.

Generating long stationary time series from experimental systems can be challenging as it demands controlling a number of factors for extended periods. The present study provides a suitable alternative by using multiple short time series realizations, hence expected to find wide applications across a number of settings. While the results presented in this study investigated the performance of a simple RNN with 10-20 neurons and a single hidden layer, the RNN hyperparameters in general will have to be tuned. The results presented showed a marked increase in accuracy across the dynamical nonlinearities generated from nonlinear deterministic processes in chaotic regimes. However, it is important to note that dynamical nonlinearities can arise across deterministic as well as non-deterministic settings. The latter would include deterministic dynamical systems with dynamical and measurement noise. Therefore, conclusions on the presence of dynamical nonlinearities do not necessarily imply presence of deterministic chaos.

**Methods**
**(a) Working Principle of the IAAFT Algorithm**
The IAAFT algorithm[28] is an iterative procedure that aims to retain the power-spectrum as well as the distribution of the given empirical sample in the surrogate realizations. As noted earlier, retaining the power-spectrum retains the linear characteristics of the time series. Rank ordering aspect of IAAFT is useful in retaining static, invertible nonlinearities but not the dynamical nonlinearities in the given empirical sample. The working principle of IAAFT is enclosed below for completeness, a detailed explanation and implementation can be found in the following references[24,27,28,34,40].

Let the given empirical sample be $\{x_n\}$.
**Step 1:** Generate a random shuffle $\{x_n^i\}$ of the given empirical sample $\{x_n\}$.
    **Step 2:** *Preserving the power spectrum in the surrogate*
    Generate the Fourier transform of $\{x_n\}$ and $\{x_n^i\}$. Let the corresponding squared amplitudes be $\{S_k^2\}$ and $\{S_k^{2i}\}$ respectively. Substitute $\{S_k^{2i}\}$ by $\{S_k^2\}$ and generate the inverse Fourier transform to obtain $\{y_n\}$.
    **Step 3:** *Preserving the distribution in the surrogate*
    Rank order $\{y_n\}$ to have same distribution as $\{x_n\}$ resulting in the surrogate $\{x_n^{i+1}\}$.
**Step 4:** Repeat Steps 2 and 3 so as to minimize the discrepancy in the spectrum between empirical sample and its surrogate.

**(b) Nonlinear Deterministic Process**
Time series were generated from discrete and continuous dynamical systems in chaotic regimes. Representative time series in chaotic regimes is shown in **Fig. 5**. Time series data for the continuous dynamical systems were generated using explicit Runge-Kutta (4, 5) implemented as a part of the MATLAB ode45 function[41].

*(i) Logistic map in chaotic regime* $(r = 4.0)^{42}$
$$x_{t+1} = rx_t(1 - x_t)$$
*(ii) Henon map in chaotic regime* $(\alpha = 1.4, \beta = 0.3)^{43,44}$
$$x_{t+1} = 1 - \alpha x_t^2 + y_t$$
$$y_{t+1} = \beta x_t$$

*(iii) Lorenz system in chaotic regime* $(\sigma = 10, \rho = 28, \beta = 8/3)^{45}$
$$\frac{dx}{dt} = \sigma(y - x)$$
$$\frac{dy}{dt} = x(\rho - z) - y$$
$$\frac{dz}{dt} = xy - \beta z$$

*(iv) Rossler system in chaotic regime* $(\alpha = 0.2, \beta = 0.2, \gamma = 5.7)^{46}$
$$\frac{dx}{dt} = -y - z$$
$$\frac{dy}{dt} = x + \alpha y$$
$$\frac{dz}{dt} = \beta + z(x - \gamma)$$

**(c) Experimental Time Series Data**

*(i) Chua's Circuit*

Chua's circuit [2,47] is a simple autonomous electric circuit and can be readily designed using resistors, capacitors, inductors and a nonlinear element. It is perhaps one of the most popular experimental evidence of deterministic chaos. An equivalent dimensionless model with parameters $(\alpha = 15.6, \beta = 28, m_0 = -8/7, m_1 = -5/7)$ has also been proposed in literature to capture the behavior of the original circuit[2,47].
$$\frac{dx}{dt} = \alpha(y - x - f(x))$$
$$\frac{dy}{dt} = x - y + z$$
$$\frac{dz}{dt} = -\beta y$$
where the piece-wise linear function $f(x) = m_1 x + 0.5(m_0 - m_1)(|x + 1| - |x - 1|)$.

*(ii) Santa-Fe Laser Time Series*

Several studies have provided compelling evidence of chaos across distinct laser systems[48-50]. The present study re-investigates Santa Fe Laser time series of 1000 samples derived from a Far-Infrared (FIR) laser in chaotic regime[51,52].

*(iii) Epileptic Seizure Time Series*

Electroencephalograms (EEG) signals recorded during epileptic seizure have been argued to exhibit patterns characteristic of nonlinear dynamical processes. Three representative EEG samples from seizure subjects reported in a recent study[3] were re-investigated using the proposed classification framework. As recommended in the original study[3], the three EEG signals were pre-processed using a 4th order low-pass Butterworth filter[1] to minimize the impact of noise and impose the high-frequency cut-off at 40Hz. In order to capture a few cycles of the EEG waveform only samples with length (N = 128) were investigated.

**(d) Nonlinear Transform of Linearly Correlated Noise**

$$x_t = \alpha x_{t-1} + \epsilon_t; \quad y_t = x_t\sqrt{|x_t|};$$

The above example was motivated by a recent study[28]. The process $x_t$ is a linearly correlated noise where $\epsilon_t$ is zero-mean, unit variance normally distributed uncorrelated noise with $y_t$ representing a static nonlinear transform of $x_t$. Several choices of the process parameters ($\alpha = 0.2, 0.4, 0.6, 0.8$) were investigated in the present study. Representative time series data generated by nonlinear transform of linearly correlated noise with process parameters ($\alpha = 0.2, 0.4, 0.6, 0.8$) is shown in **Fig. 3**.

**(e) Surrogate Testing Using a Recurrent Neural Network**

*Data:* The time series realizations was fixed at (N = 1000) across all the data sets. Time series of three different lengths (L = 32, 64, 128) were investigated. For the experimental data sets in the present study, (N = 1000) realizations was generated by randomly choosing a sequence of time points of length (L = 32, 64, 128) from the given data. Representative samples of the various time series are shown in **Fig. 3**.

*RNN:* RNN architectures by very design are ideal for prediction and classification of sequence data. RNN cell unfolded in time[53,54] and a typical RNN architecture comprising of multiple RNN cells in the hidden layer is shown in **Fig. 2**. In the present study, the input and output of the RNN were the time series realizations and their corresponding labels respectively. The time series realizations (N = 1000) was split into training samples (75%) and test samples (25%). Since each time series realization was paired to its IAAFT surrogate counterpart, the classes were balanced by very design justifying the choice of accuracy as a classifier performance measure in the present study. RNN parameters were chosen after experimentation[55]. RNN was implemented using Keras high-level neural network API with Tensorflow backend[53,54] and Adam optimizer (ADAM)[56] (learning rate 0.0001, batch size 16 and binary cross-entropy loss) for the data sets in the present study. The number of neurons for the synthetic data sets generated from nonlinear dynamical systems, was chosen as (N = 10), **Table 1**. For the nonlinearly correlated noise, the number of neurons was also fixed at (N = 10), **Table 1**. For the experimental time series data, the number of hidden neurons varied and enclosed in **Table 1**. Neurons in the hidden layer were accompanied by rectified linear unit (ReLU) activation function whereas those in the output layer had sigmoid activation function. RNN learning curves were inspected during the training phase for potential overfitting. The validation split in the training phase was set at 30%, implying the last 30% of the training data were used as internal validation in computing the accuracy and loss curves as a function of the epoch. The training and validation loss as a function of the epoch for representative nonlinear deterministic processes and experimental time series are shown in **Fig. 4**. As can be observed for each of these cases, the training and validation loss simultaneously transitioned to markedly lower values with increasing epochs. While certain RNN applications do encourage having a validation loss lower than that of the training loss, the present study estimated the accuracies (**Table I**) at the epoch where the training and validation loss were simultaneously low, **Fig. 4**. A smoothing window of five samples was used to generate the learning curves, **Fig. 4**, and accuracy profiles, **Figs. 5-7**, as a function of the epochs.

**Figure 1**

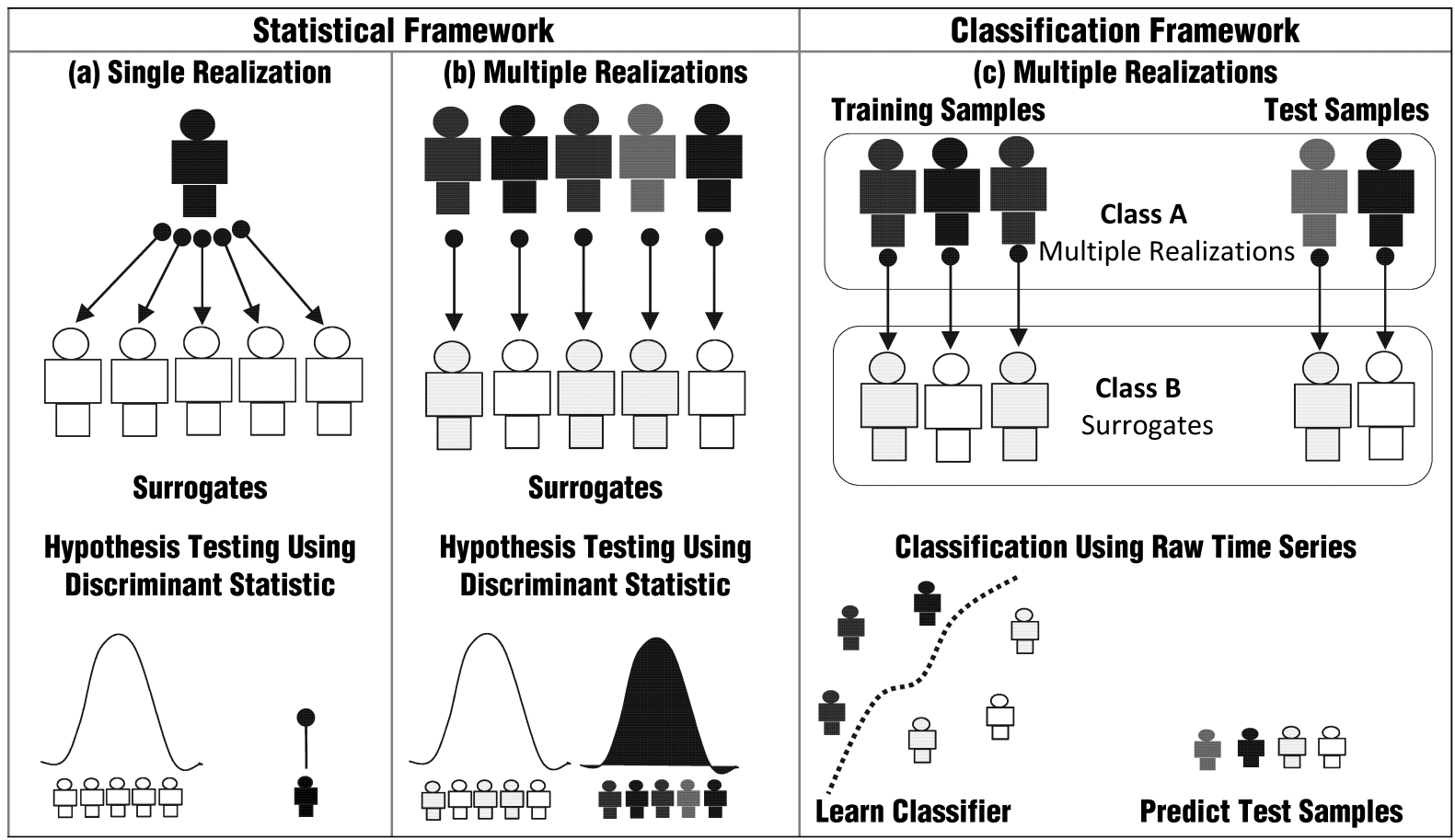

Figure 2

(a) **RNN Cell**

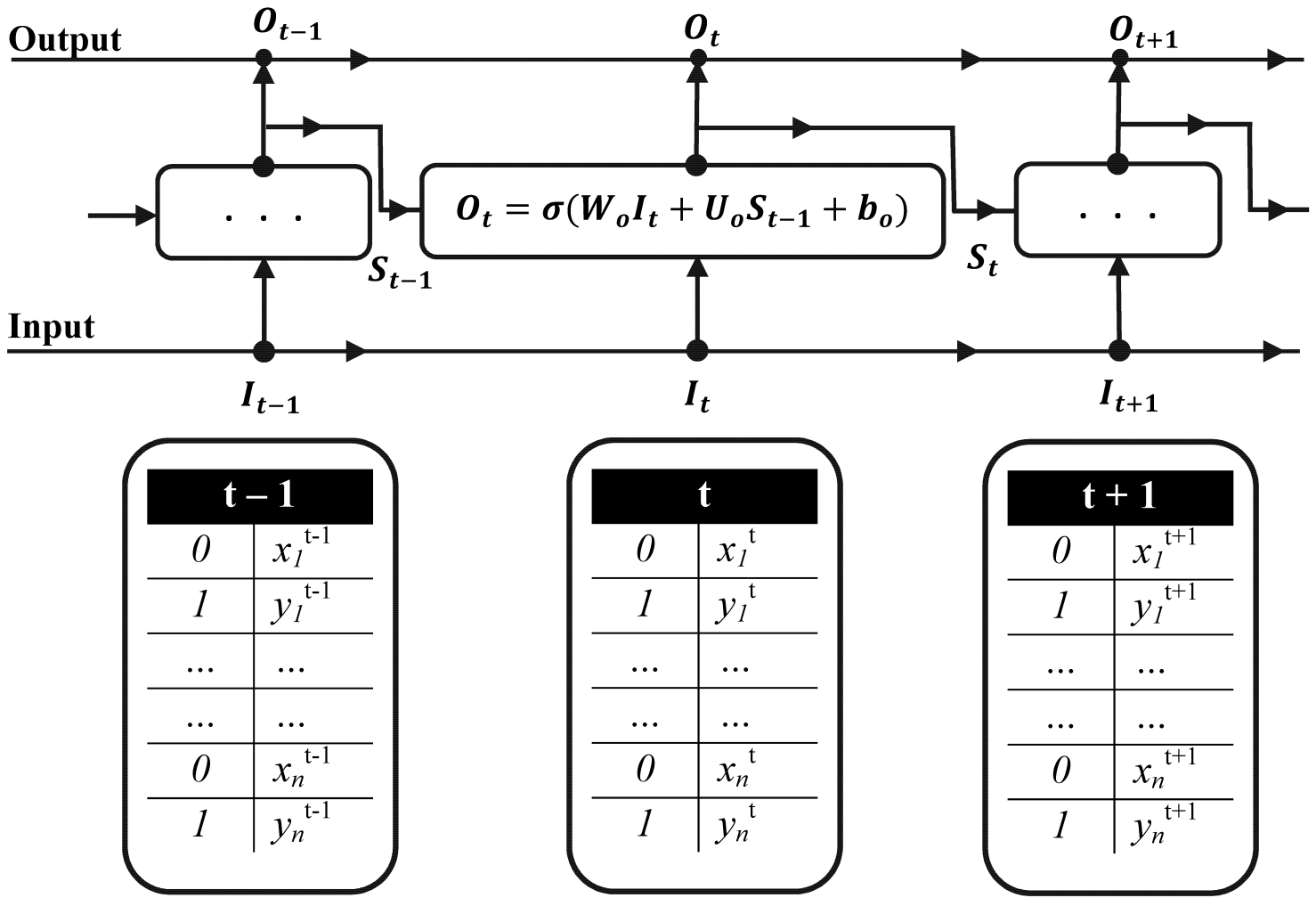

(b) **RNN Architecture**

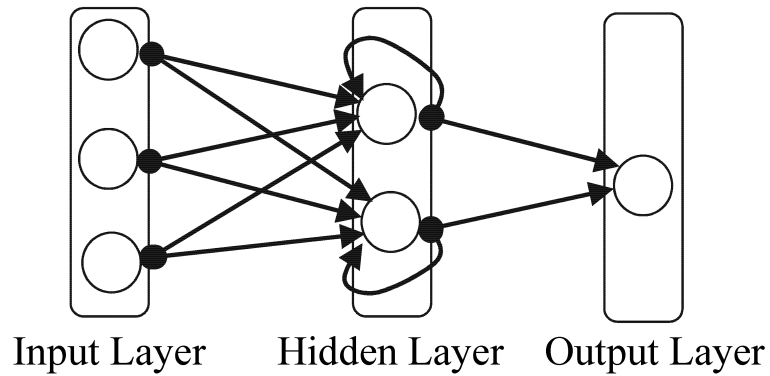

**Figure 3**

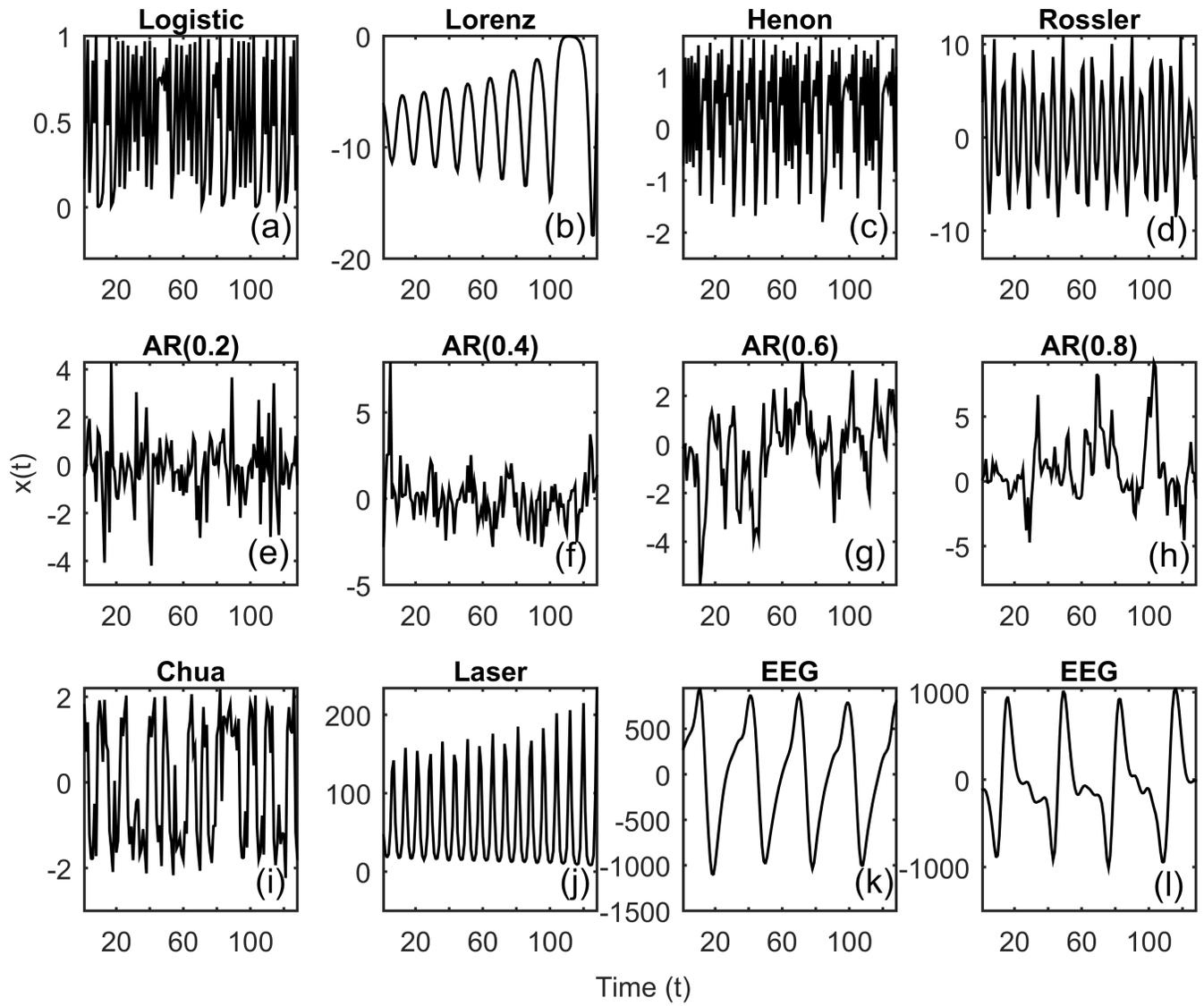

**Figure 4**

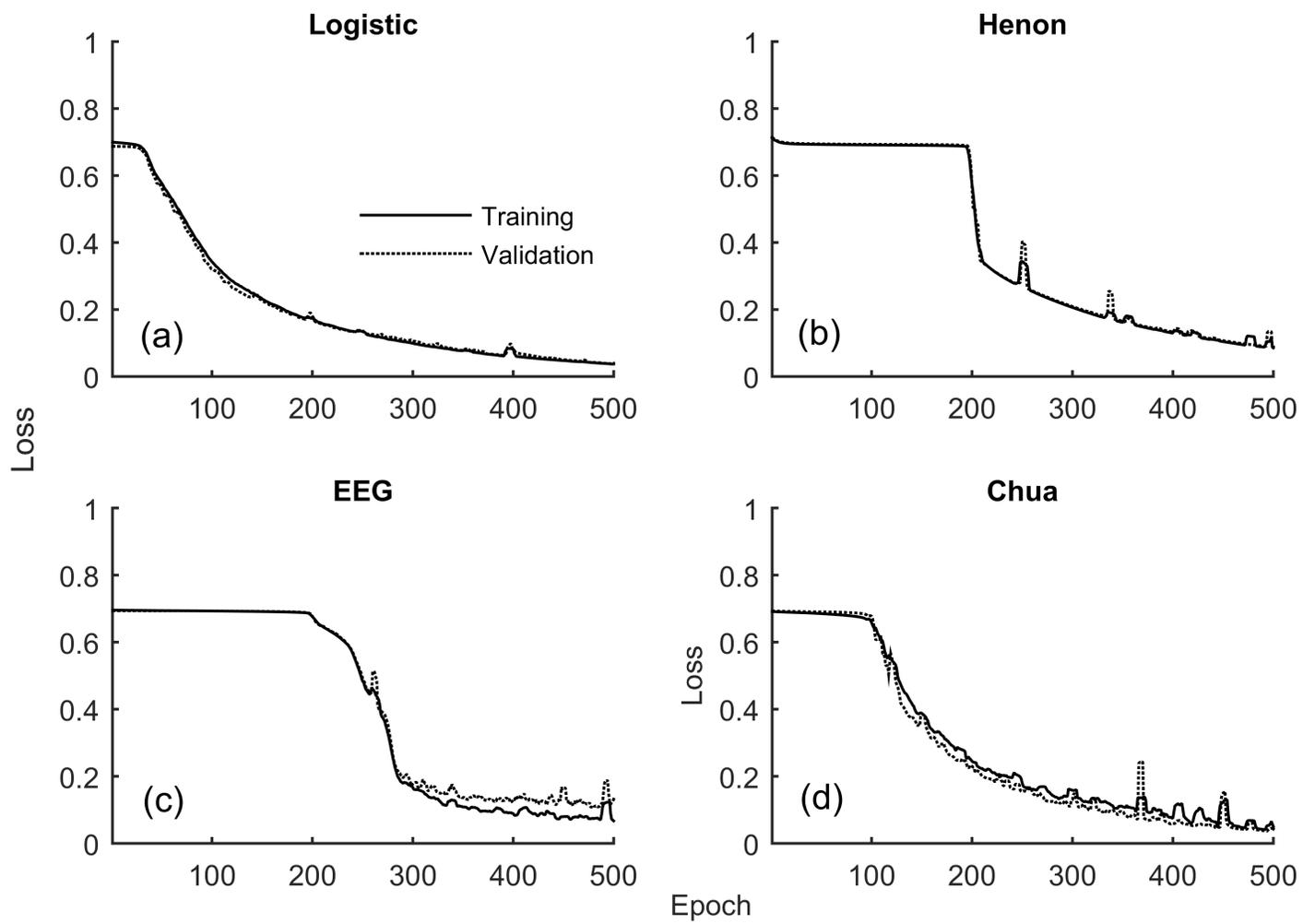

Figure 5

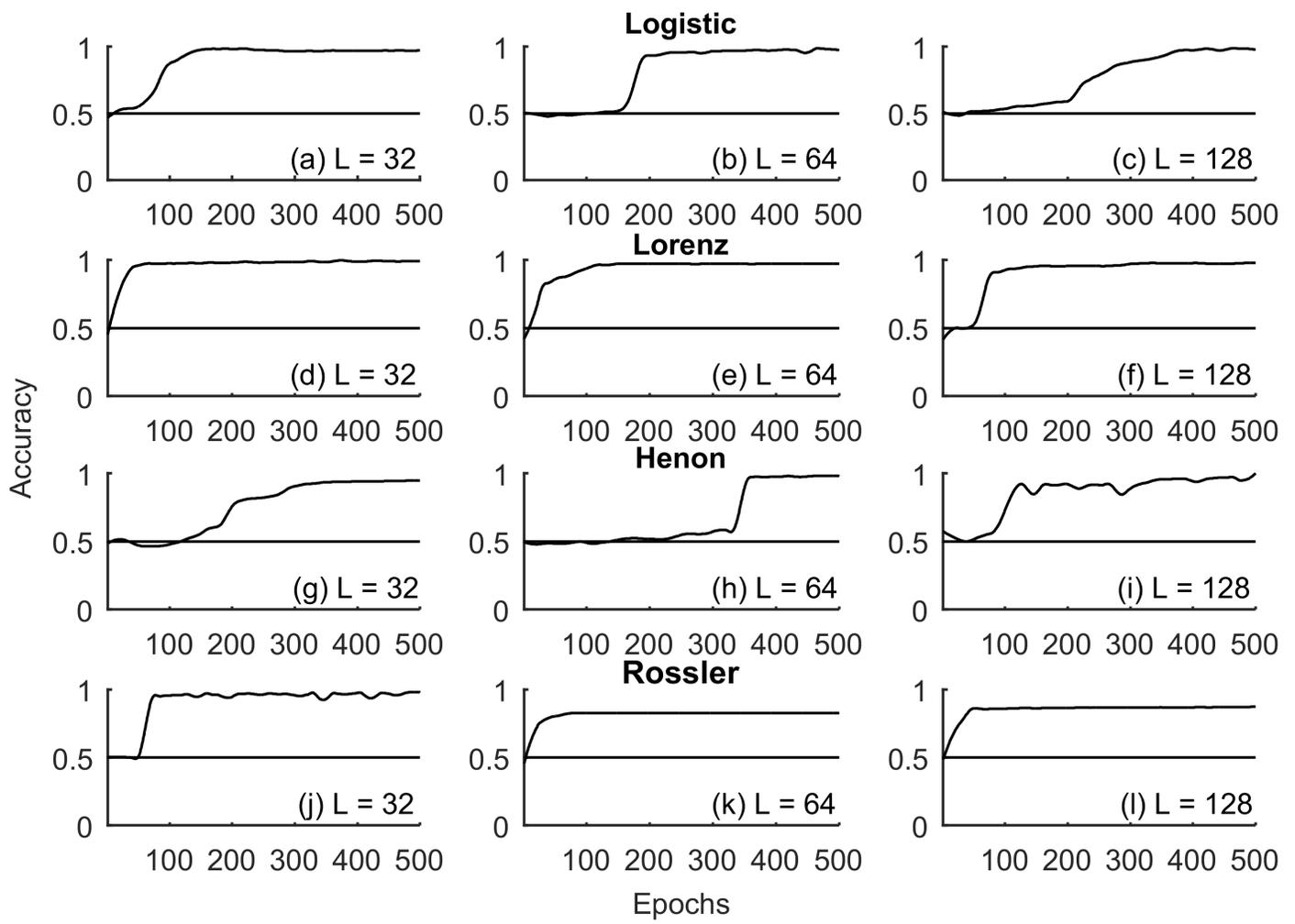

**Figure 6**

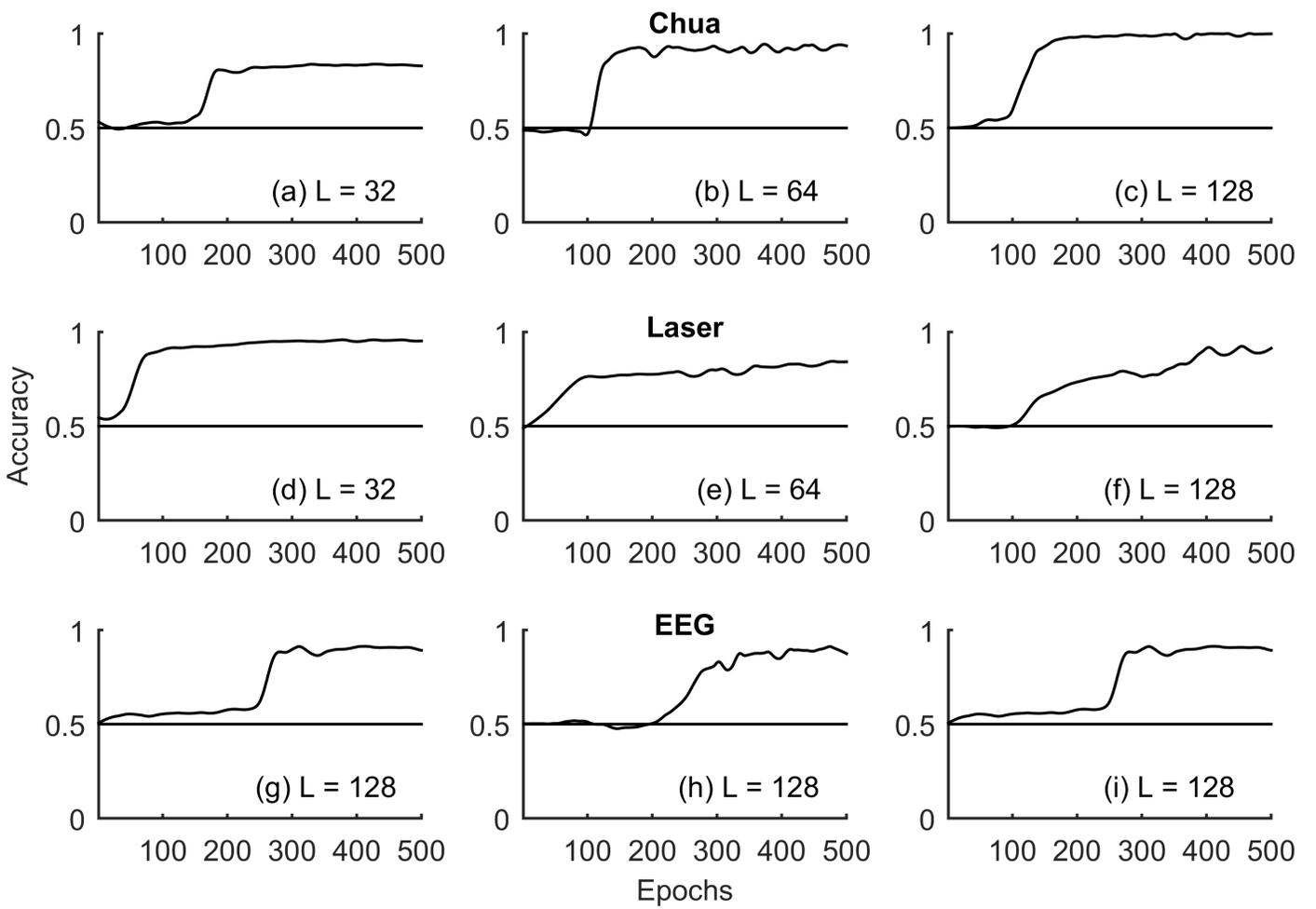

**Figure 7**

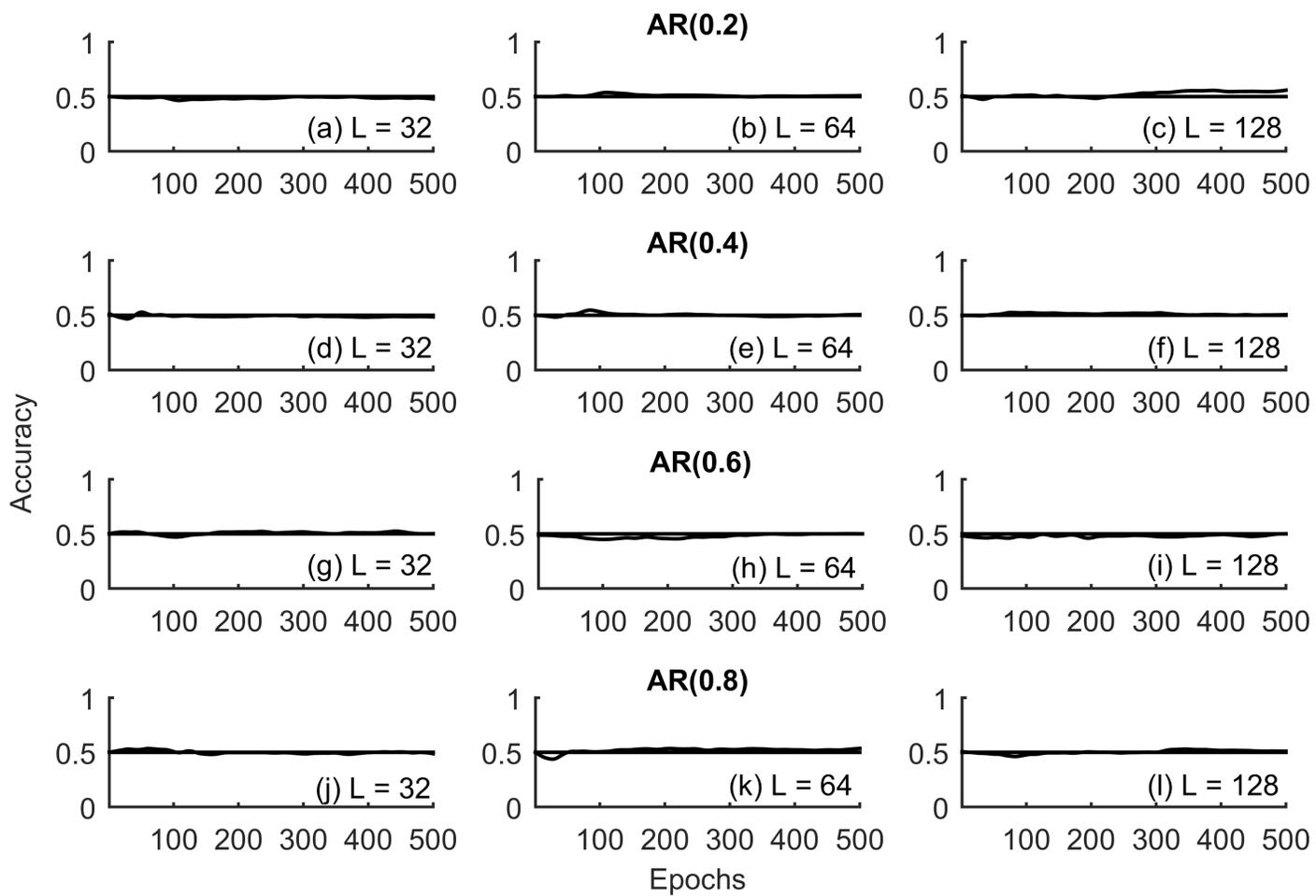